\documentclass[aps,prb,reprint,superscriptaddress,showpacs]{revtex4-1}

\usepackage{graphicx}
\usepackage{nicefrac}
\usepackage{color}
\usepackage{braket}
\usepackage{tabularx}
\begin{document}

\title{Crystal structure and phonon softening in Ca$_3$Ir$_4$Sn$_{13}$}

\author{D.~G.~Mazzone}
\altaffiliation{daniel.mazzone@psi.ch}
\affiliation{Laboratory for Neutron Scattering and Imaging, Paul Scherrer Institut, CH-5232 Villigen, Switzerland}
\author{S.~Gerber}
\altaffiliation{Present address: Stanford Institute for Materials and Energy Sciences, SLAC National Accelerator Laboratory, Menlo Park, California 94025, USA}
\affiliation{Laboratory for Neutron Scattering and Imaging, Paul Scherrer Institut, CH-5232 Villigen, Switzerland}
\author{J.~L.~Gavilano}
\affiliation{Laboratory for Neutron Scattering and Imaging, Paul Scherrer Institut, CH-5232 Villigen, Switzerland}
\author{R.~Sibille}
\affiliation{Laboratory for Developments and Methods, Paul Scherrer Institut, CH-5232 Villigen, Switzerland}
\author{M.~Medarde}
\affiliation{Laboratory for Developments and Methods, Paul Scherrer Institut, CH-5232 Villigen, Switzerland}
\author{B.~Delley}
\affiliation{Condensed Matter Theory Group, Paul Scherrer Institut, CH-5232 Villigen PSI, Switzerland}
\author{M.~Ramakrishnan}
\altaffiliation{Present address: Swiss Light Source, Paul Scherrer Institut, CH-5232 Villigen PSI, Switzerland}
\affiliation{Laboratory for Neutron Scattering and Imaging, Paul Scherrer Institut, CH-5232 Villigen, Switzerland}
\author{M.~Neugebauer}
\affiliation{Laboratory for Neutron Scattering and Imaging, Paul Scherrer Institut, CH-5232 Villigen, Switzerland}
\author{L. P.~Regnault}
\affiliation{SPSMS, UMR-E CEA/UJF-Grenoble 1, INAC, F-38054 Grenoble, France}
\author{D.~Chernyshov}
\affiliation{Swiss-Norwegian Beamlines, European Synchrotron Radiation Facility, BP220, F-38043 Grenoble, France}
\author{A.~Piovano}
\affiliation{Institut Laue-Langevin, F-38043 Grenoble, France}
\author{T.~M.~Fern\'andez-D\'iaz}
\affiliation{Institut Laue-Langevin, F-38043 Grenoble, France}
\author{L.~Keller}
\affiliation{Laboratory for Neutron Scattering and Imaging, Paul Scherrer Institut, CH-5232 Villigen, Switzerland}
\author{A.~Cervellino}
\affiliation{Swiss Light Source, Paul Scherrer Institut, CH-5232 Villigen PSI, Switzerland}
\author{E.~Pomjakushina}
\affiliation{Laboratory for Developments and Methods, Paul Scherrer Institut, CH-5232 Villigen, Switzerland}
\author{K.~Conder}
\affiliation{Laboratory for Developments and Methods, Paul Scherrer Institut, CH-5232 Villigen, Switzerland}
\author{M.~Kenzelmann}
\affiliation{Laboratory for Developments and Methods, Paul Scherrer Institut, CH-5232 Villigen, Switzerland}
\date{\today}

\begin{abstract} 
We investigated the crystal structure and lattice excitations of the ternary intermetallic stannide Ca$_3$Ir$_4$Sn$_{13}$ using neutron and x-ray scattering techniques.  For $T$ $>$ $T^*$~$\approx$~38~K the x-ray diffraction data can be satisfactorily refined using the space group $Pm\bar{3}n$. Below  $T^*$ the crystal structure is modulated with a propagation vector of $\vec{q}$~=~(1/2,~1/2,~0). This may arise from a merohedral twinning in which three tetragonal domains overlap to mimic a higher symmetry, or from a doubling of the cubic unit cell. Neutron diffraction and neutron spectroscopy results show that the structural transition at $T^*$ is of a second-order, and that it is well described by mean-field theory. Inelastic neutron scattering data point towards a displacive structural transition at $T^*$ arising from the softening of a low-energy phonon mode with an energy gap of $\Delta$(120~K)~=~1.05~meV. Using density functional theory the soft phonon mode is identified as a 'breathing' mode of the Sn$_{12}$ icosahedra and is consistent with the thermal ellipsoids of the Sn2 atoms found by single crystal diffraction data.
\end{abstract}


\maketitle 

\section{Introduction} 

The family of compounds with the general formula $R_3M_4$Sn$_{13}$, where $R$ is an alkali metal or a rare earth and $M$ is a transition metal (Ir, Rh or Co), was first synthesized in 1980 \cite{Remeika1980, Espinosa1980, Cooper1980}. Recently, it regained interest among the condensed matter community due to its unusual behavior at low temperatures, which is characterized by the existence of structural phase transitions, a change of the Fermi surface and superconductivity\cite{Yang2010, Wang2012, Gerber2013, Biswas2014, Klintberg2012, Tompsett2014, Fang2014, Goh2011, Zho2012, Kase2011, Hayamizu2011, Hayamizu2010, Kumar2014, Kumar20142, Kuo2014,Mardegan2013, Levett2003, Mazzone2014}.

Ca$_3$Ir$_4$Sn$_{13}$ is a good model material to study the physics of this family. It displays an anomaly in the resistivity at $T^*$~$ \approx$~38~K and a superconducting transition at $T_c$~$\approx$~7~K \cite{Espinosa1980, Cooper1980, Yang2010, Wang2012}. Macroscopic transport measurements\cite{Yang2010} attributed the anomaly at $T^*$ to ferromagnetic spin fluctuations close to the Fermi level, hinting at superconductivity mediated by ferromagnetic fluctuations. However, muon spin rotation\cite{Gerber2013, Biswas2014} later showed that magnetic fluctuations play no role on the physical properties of (Ca,~Sr)$_3$Ir$_4$Sn$_{13}$. The muon measurements also pointed towards a nodeless superconducting order parameter which is well described by BCS theory in the strong-coupling limit with a London penetration depth\cite{Gerber2013} $\lambda_L$~$\approx$~385~nm. 

Based on transport measurements\cite{Wang2012} including Hall and Seebeck coefficients, an alternative scenario was proposed that involves the onset of a charge-density wave (CDW) at $T^*$. Further support for this mechanism was recently provided by optical spectroscopy measurements\cite{Fang2014}, which suggested the existence of an unconventional CDW in the closely related compound Sr$_3$Ir$_4$Sn$_{13}$. In the low-temperature phase, Fang $et.$ $al.$\cite{Fang2014} reported a suppression of the optical conductivity, consistent with an energy gap in the spectrum of electronic excitations. However, this suppression appears at a much higher energy scale than expected for conventional CDW transitions.

In the case of Ca$_3$Ir$_4$Sn$_{13}$ the pressure-temperature phase diagram reveals a dome-shape superconducting phase\cite{Klintberg2012, Goh2011} with a broad maximum at $T$~=~8.9~K and $p$~=~40~kbar. Concomitantly, a rapid decrease of $T^*$ was observed with increasing pressure, suggesting a pressure-induced structural quantum critical point.

For Sr$_3$Ir$_4$Sn$_{13}$ it was claimed\cite{Klintberg2012} that the structural transition at $T^*$~$\approx$~147 K converts the simple cubic high-temperature structure ($Pm\bar{3}n$) into a body-centered cubic structure with twice the lattice constant ($I\bar{4}3d$). Whereas this would correspond to simultaneous ordering along the face centered cubic reciprocal vectors $\vec{q}$~=~(1/2,~1/2,~0), (1/2,~0,~1/2), (0,~1/2,~1/2), Lindhard function calculations\cite{Klintberg2012} suggest $\vec{q}$~=~(1/2,~1/2,~1/2) as a possible ordering wavevector, and  subsequent band structure calculations\cite{Tompsett2014} revealed phonon instabilities at $\vec{q}$~=~(1/2,~0,~0) and (1/2,~1/2,~0) for (Ca,~Sr)$_3$Ir$_4$Sn$_{13}$. Therefore, the question of the propagation vector describing the structural modulation of the low-temperature phase as well as the the microscopic mechanism at the origin of the transition at $T^*$ remains open.
 
\begin{figure}[h] 
	\includegraphics[width=\linewidth]{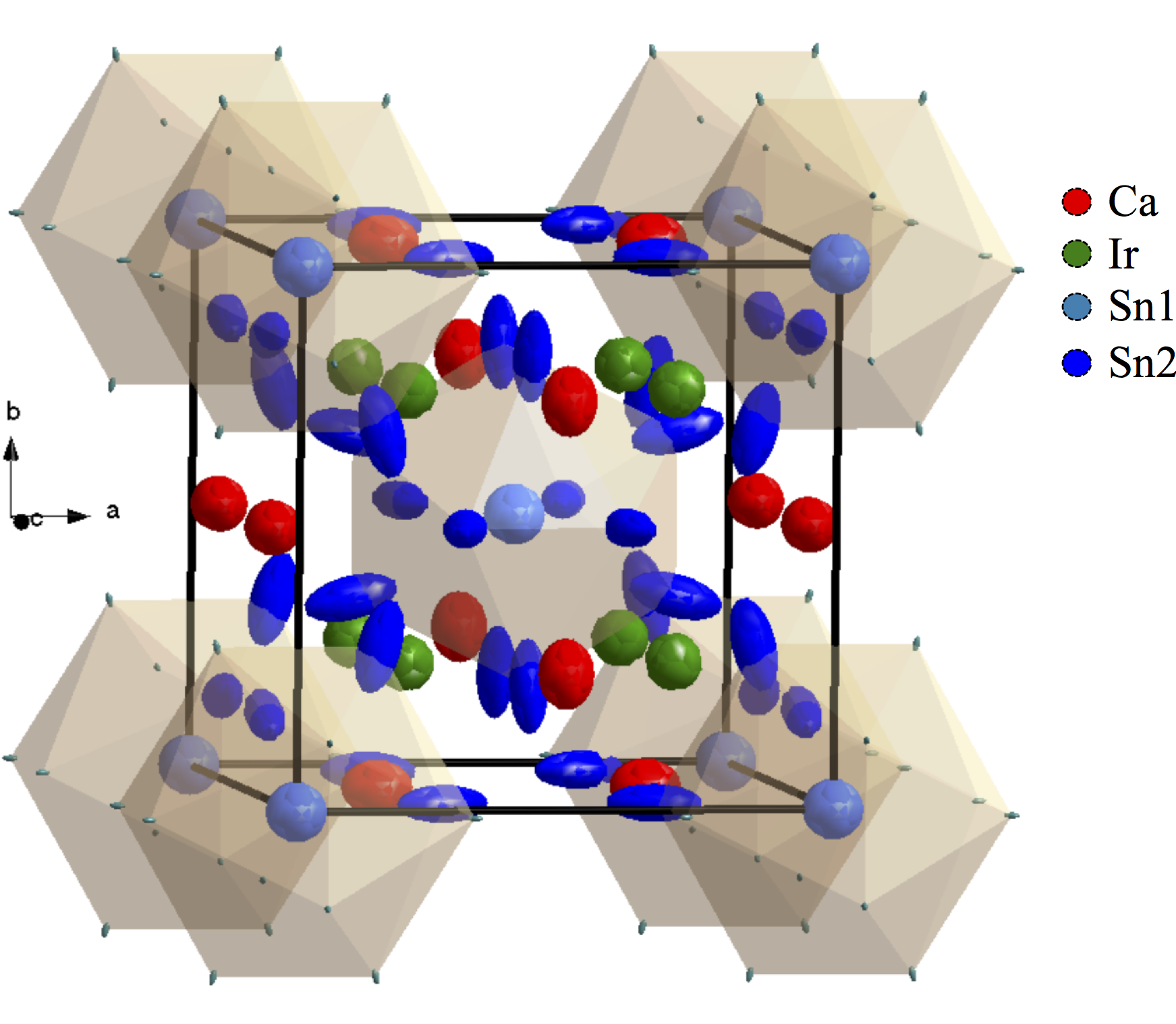}
	\caption{\label{crystalstructure}
		(Color online) Crystal structure of Ca$_3$Ir$_4$Sn$_{13}$ measured by single crystal neutron diffraction and refined in $Pm\bar{3}n$ for $T$~$>$~$T^*$. Color code: red Ca, green Ir, blue Sn. The plotted ellipsoids denote the anisotropic mean-square displacements of the atoms (with a magnification factor of two for clarity).}
\end{figure}

Here, we studied the crystal structure and lattice excitations of Ca$_3$Ir$_4$Sn$_{13}$ above and below $T^*$ using x-ray and neutron diffraction as well as neutron spectroscopy. Above $T^*$ the data can be satisfactorily refined using the cubic space group $Pm\bar{3}n$ in agreement with earlier reports \cite{Cooper1980}. The microscopic investigations reveal a low-temperature superstructure with a propagation vector of $\vec{q}$~=~(1/2,~1/2,~0). The unsatisfactory refinement of the x-ray diffraction data in the $I$-centered space groups speaks against a doubling of the cubic unit cell along the three directions. Neutron diffraction and spectroscopy data show that the phase transition at $T^*$ is of a second-order. Moreover, inelastic neutron scattering data reveal a softening of a low-energy phonon above $T^*$ suggesting a displacive phase transition. A freezing of a 'breathing' mode of the Sn$_{12}$ icosahedra in (Ca,~Sr)$_3$Ir$_4$Sn$_{13}$ as suggested by density functional theory, is consistent with the diffraction data. The atomic displacements associated with this soft mode are in excellent agreement with the shape of the thermal ellipsoids of the Sn2 atoms in the Sn$_{12}$ icosahedra of Ca$_3$Ir$_4$Sn$_{13}$ (see Fig. \ref{crystalstructure} ).
\section{Results and Discussion} 
\subsection{Experimental Details}
A Sn self-flux method was used to synthesize gram-sized high-quality single crystals. High purity elements (with atomic parts of 3 Ca, 4 Ir and a Sn flux) were heated to 1050~$^\circ$C in an evacuated and sealed quartz tube. After two hours, the liquid was cooled to 520~$^\circ$C with a cooling rate of 4~$^\circ$C/h and then quenched to room temperature. The excess flux was removed with HCl.

X-ray powder diffraction experiments were carried out at the Material Science beamline of the Swiss Light Source at the Paul Scherrer Institut, Villigen, Switzerland. Single crystal x-ray diffraction experiments were conducted on the single crystal x-ray diffractometer BM01A of the Swiss-Norwegian Beamlines at the European Synchrotron Radiation Facility, Grenoble, France in a temperature range from $T$~=~5 to 280 K. We prepared a  needle shaped, sub-millimeter sized single-crystalline sample, which was exposed to a x-ray beam of $\lambda$~=~0.694~\AA~perpendicular to the main axis of the crystal. For the data reduction CrysAlisPRO\cite{crysalis} was used and the structure was refined by full-matrix least squares on $F^2$ using $SHELX-97$\cite{shellx}. 

Neutron powder diffraction measurements were carried out on DMC  at the Swiss Spallation Neutron Source SINQ at the Paul Scherrer Institut, Villigen, Switzerland.
Single crystal neutron scattering results are based on experiments performed on the triple-axis spectrometers IN22, IN8 and the single crystal diffractometer D9 at the Institut Laue-Langevin, Grenoble, France. For the measurements carried out on IN22 and IN8 we used a sample consisting of two single crystals of 1.6 and 0.9~g, which were coaligned with an accuracy of 0.5~$^\circ$ using the single crystal diffractometer Morpheus at the Swiss Spallation Neutron Source SINQ at the Paul Scherrer Institut, Villigen, Switzerland. Diffraction studies on IN22 were performed between 1.8 and 50 K fixing the analyzer of the instrument ($k_f$~=~2.662~\AA$^{-1}$). Inelastic neutron scattering results, conducted on IN22, were performed at $k_f$~=~1.97~\AA$^{-1}$. We studied the excitation spectra around the superstructure satellite $\vec{Q}$~=~(7/2,~7/2,~1) for energy transfers $\Delta E$~=~0.7~-~2.8~meV and temperatures between $T$~=~20 and 120 K. Energy scans at each temperature were fitted by means of two Gaussians and a constant background, describing the elastic incoherent contribution and one phonon at finite energy. The motion of the atoms was experimentally studied by means of single crystal neutron diffraction on D9. A single crystal (3.5~x~3.5~x~4.3~mm$^3$) was exposed to a neutron wavelength of $\lambda$~=~0.84~\AA. Integrated intensities from data collected at $T$~=~50~K were refined in the space group $Pm\bar{3}n$ using the program $FULLPROF$\cite{Fullprof}.
\subsection{Structural studies} 

\begin{figure}[h] 
	\includegraphics[width=\linewidth]{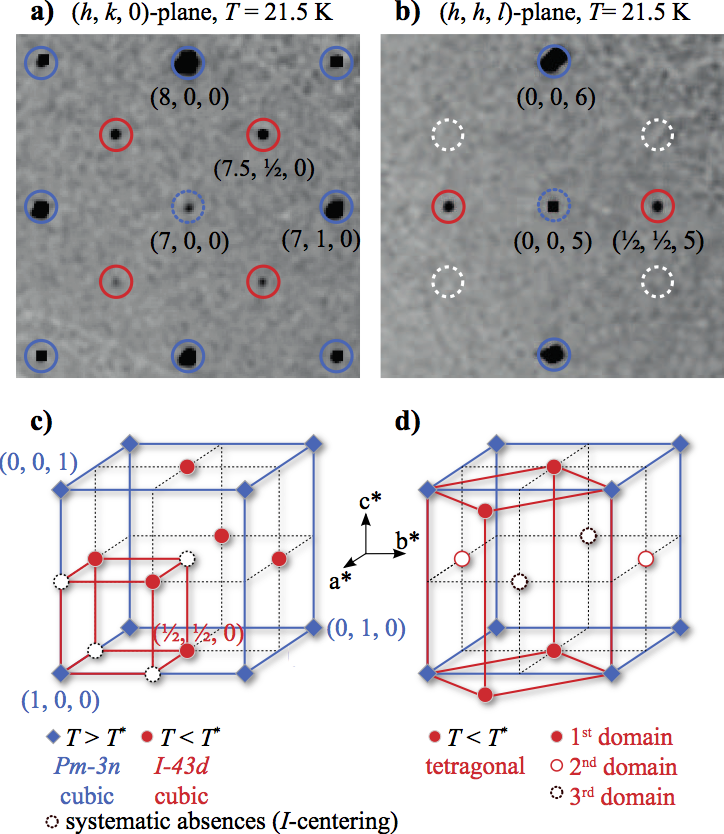}
	\caption{\label{xraydiff}
		(Color online) Two cuts of the reciprocal space of Ca$_3$Ir$_4$Sn$_{13}$: \textbf{a)} the ($h$,~$k$,~0)- and \textbf{b)} the ($h$,~$h$,~$l$)-plane as measured on BM01A at  $T$~=~21.5~K. The ($h$,~$k$,~$l$) indexing corresponds to the cubic $Pm\bar{3}n$ lattice. The fundamental reflections originating from the space group $Pm\bar{3}n$ are emphasized in blue. Additional (superstructure) reflections are represented as red circles and can be indexed with the propagation vector $\vec{q}$~=~(1/2,~1/2,~0). Bottom: reciprocal lattice for \textbf{c)} the double-cubic unit cell description proposed for Sr$_3$Ir$_4$Sn$_{13}$\cite{Klintberg2012} and \textbf{d)} the alternative description with three tetragonal domains.}
\end{figure} 

The collected single crystal x-ray diffraction data at $T$~$>$~$T^*$ were satisfactorily refined using the cubic space group $Pm\bar{3}n$. The conventional agreement factor of the fit, $R_1$, equals 1.5~\%. The data reveal a lattice constant of $a$~=~9.70880(4)~\AA~and $a$~=~9.6892(1)~\AA~for $T$~=~280 and 80 K, respectively. Between $T$~=~280 and 40~K the lattice parameter decreases linearly, reflecting the thermal contraction. Our results agree with previous reports of the high-temperature crystal structure\cite{Cooper1980}.

The diffraction patterns reveal additional Bragg reflections below $T^*$ at $\vec{Q}$~=~$\vec{k}$~$\pm$~$\vec{q}$ (see red circles in Fig. \ref{xraydiff} a and b), with $\vec{k}$ a wavevector of the high-temperature phase and $\vec{q}$~=~\{(1/2,~1/2,~0), (1/2,~0,~1/2), (0,~1/2,~1/2)\}. Thus, the observed systematic absences are incompatible with a propagation vector $\vec{q}$~=~(1/2,~1/2,~1/2), in agreement with band structure calculations \cite{Tompsett2014}. This is illustrated in Fig. \ref{xraydiff} b, where the white circles represent the missing reflections associated with a modulation $\vec{q}$~=~\{(1/2,~1/2,~1/2)\}.

Our experiment is evidence for either a $I$-centered double-cubic unit cell (in that case the missing reflections on Fig. \ref{xraydiff} c are forbidden by the body-centered lattice) or the formation of three equivalent tetragonal domains (merohedral twinning that mimic a higher symmetry). In the latter case each of the tetragonal axes would be oriented along one of the high-temperature cubic axes (see Fig. \ref{xraydiff} d). Attempts to refine the crystal structure below $T^*$ in a $I$-centered cubic unit cell were not satisfactory ($R_1$~$\sim$~6~\%).

A complete interpretation of the diffraction data of Ca$_3$Ir$_4$Sn$_{13}$ below $T^*$ is beyond the scope of this manuscript and will be published separately.

\subsection{Characterization of the phase transition at $T^*$}

\begin{figure}[h] 
	\includegraphics[width=\linewidth]{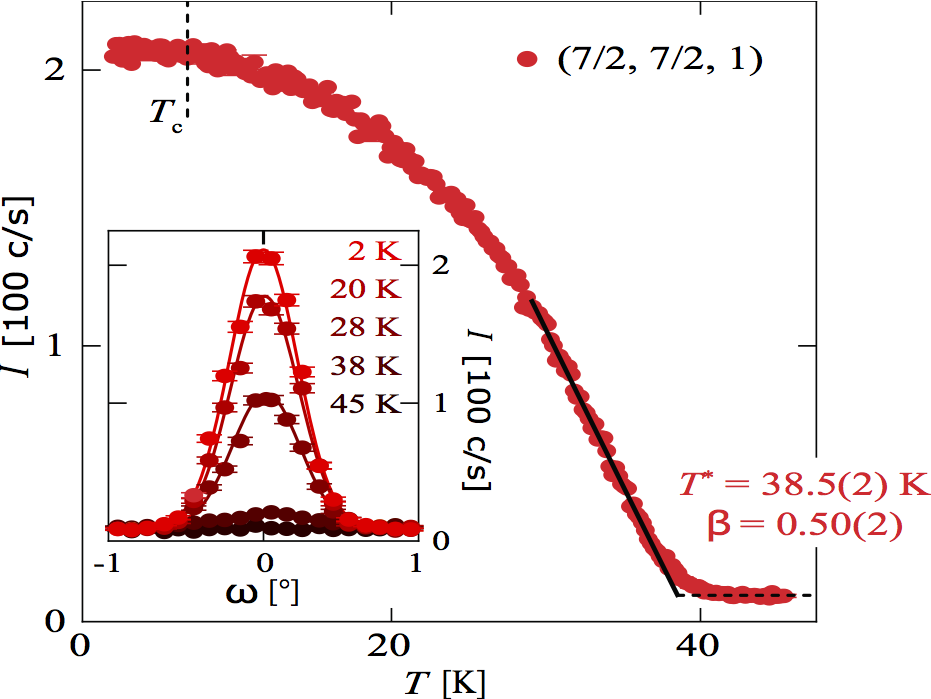}
	\caption{\label{In22diff}
		(Color online) Temperature dependence of the superstructure Bragg reflection at $\vec{Q}$~=~(7/2,~7/2,~1) with a critical exponent of $\beta$~=~0.50(2) and $T^*$~=~38.5(2) K. The inset shows diffraction patterns for $T$~$<$~$T^*$ measured at selected temperatures.}
\end{figure}

The nature of the phase transition was investigated on the instrument IN22 around the superstructure satellite $\vec{Q}$~=~(7/2,~7/2,~1). The inset of Fig. \ref{In22diff} depicts this superstructure satellite as a function of the sample orientation $\omega$ for selected temperatures $T$~$<$~$T^*$. Neither the position nor the full width half maximum (FWHM) $\Gamma$ is temperature dependent, so a measurement of the maximal intensity, $I$, is proportional to the integrated intensity.

Fig. \ref{In22diff} shows the maximal intensity of the superstructure satellite $\vec{Q}$~=~(7/2,~7/2,~1) as a function of temperature. We find a continuous phase transition at $T^*$. A gradual increase of nuclear intensity is observed with decreasing temperature below $T$~=~40 K. A saturation of the nuclear intensity is found only at very low temperatures, around the superconducting transition at $T_c$~$\approx$~7~K.

The temperature dependence of the arising nuclear intensity below $T$~=~40 K  was fitted by the scaling law $I$~=~$I_0$($T^*$~-~$T$)$^{2\beta}$. The best fit to the data near $T^*$ corresponds to a critical exponent of $\beta$~=~0.50(2) and a critical temperature of $T^*$~=~38.5(2) K (see black line in Fig. \ref{In22diff}). Our data thus indicate a second-order transition with a mean-field critical exponent.

\begin{figure}[h] 
	\includegraphics[width=\linewidth]{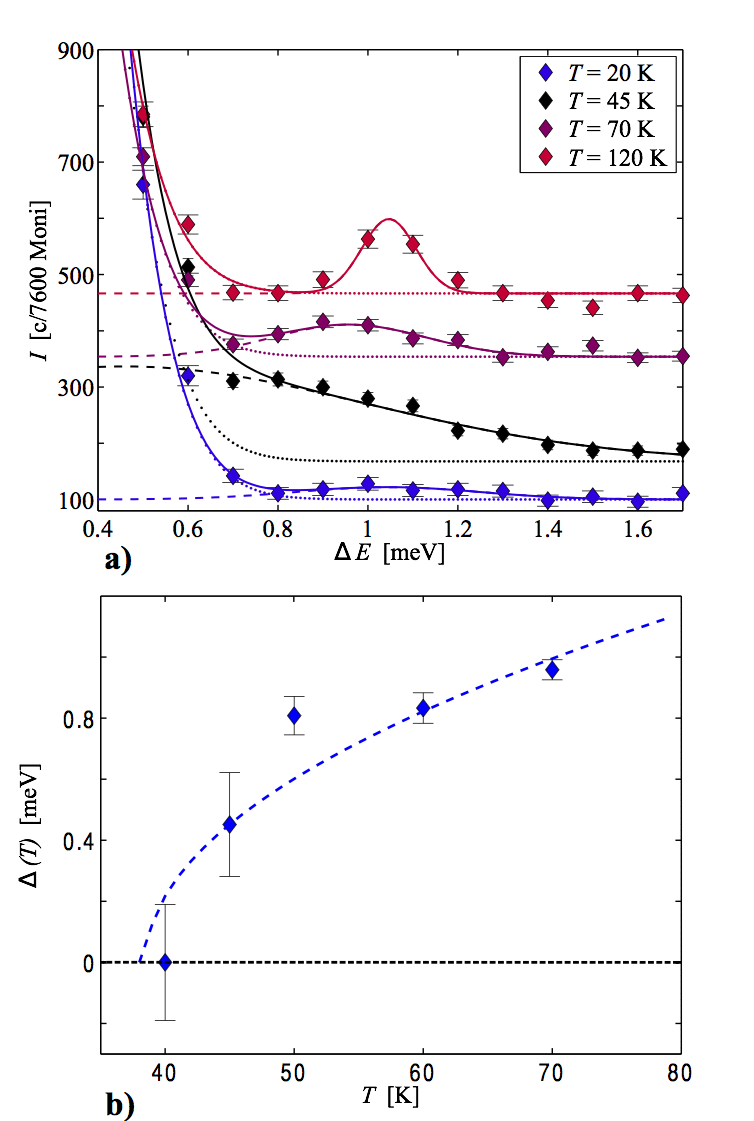}
	\caption{\label{inelasticIn22}
		(Color online) \textbf{a)} Energy scans at $\vec{Q}$~=~(7/2,~7/2,~1) and $T$~=~120, 70, 45 and 20~K (scans at different temperatures are plotted with a relative offset). The solid line denotes the total fit, the dotted line the the elastic coherent contribution and the dashed line the phonon. \textbf{b)} Softening of the $\vec{Q}$~=~(7/2,~7/2,~1) phonon above $T^*$. The dashed line correspond to a mean-field behavior $\Delta(T)$~$\propto$~($T$~-~$T^*$)$^{\beta}$ with $\beta$~=~1/2.}
\end{figure}

Further investigations of the phase transition were performed on IN22 by means of inelastic neutron scattering. Energy scans at $\vec{Q}$~=~(7/2,~7/2,~1) and $T$~=~120, 70, 45 and 20~K are shown in Fig. \ref{inelasticIn22} a. With decreasing temperature a low-energy phonon (1.05 meV at 120 K) shifts towards lower energy transfers, revealing a phonon softening towards $T^*$. While the incoherent peak and the phonon excitation are separated at temperatures higher than 60 K, they mix for $T$ $<$ 60 K. The observed phonon below $T^*$ (1.04~meV at 20~K) is reminiscent to the temperature dependent energy transfer in the structural transition\cite{Scott1974} of SrTiO$_3$. The phonon peak appears to broaden in energy close to  $T^*$. In order to obtain the excitation energy of the phonon $\Delta$ as a function of temperature we thus fixed\cite{additionalnote1} the width of the phonon to FWHM~$\propto$~($T$~-~$T^*$)$^{-1}$.

The temperature dependence of $\Delta$, shown in Fig. \ref{inelasticIn22}~b, suggest the existence of a displacive, second-order phase transition at $T^*$. We observe a gradual decrease of the phonon energy for decreasing temperatures below 70 K. We compare the temperature dependence of a softening expected in a mean-field model, $\Delta(T)$~$\propto$~($T$~-~$T^*$)$^{\beta}$ with $\beta$~=~1/2, to the observed phonon energies, and find a reasonably good agreement with the experiment. This shows that the dynamic critical behavior is consistent with that found from an order parameter analysis (see Fig. \ref{In22diff} and Fig. \ref{inelasticIn22}~b).

The results from neutron diffraction and neutron spectroscopy thus both suggest the existence of a second-order phase transition at $T^*$. The softening of a phonon and the description in a mean-field model hints for a CDW transition. However, given the complexity of the Fermi surface of this material\cite{Tompsett2014}, evidence of strong nesting at $\vec{q}$~=~(1/2,~1/2,~0) and an opening of a CDW gap has to be directly confirmed.

\subsection{Identification of the soft phonon mode} 
\subsubsection{Density functional theory calculations}
We investigated the phonon modes in the two isostructural compounds Ca$_3$Ir$_4$Sn$_{13}$ and Sr$_3$Ir$_4$Sn$_{13}$ using the PBEsol approximation \cite{Perdew} of density functional theory (DFT). The experimental lattice constant for the Sr compound is 0.88~\% larger than for the Ca, leaving more 
room for the Sn atom to rattle around in its cage. Indeed,  Klintberg $et.$ $al.$ \cite{Klintberg2012} showed that Sr$_3$Ir$_4$Sn$_{13}$ has a higher $T^*$, which indicates a more pronounced soft mode.

Phonon modes were calculated in the present work for super-cells
mapping the (1/2,~1/2,~0) zone boundary modes onto the $\Gamma$ point. Experimental lattice parameters were used with otherwise relaxed atomic positions. While Ca$_3$Ir$_4$Sn$_{13}$ in these calculations is a stable compound, we find consistently a soft mode at $\vec{q}$~=~(1/2,~1/2,~0)  for Sr$_3$Ir$_4$Sn$_{13}$. However, the calculated imaginary frequency of about 1$i$~THz $\sim$ 4~meV $\sim$ 50~K underestimates the tendency for instability, which may explain why the calculation does not reproduce the transition in Ca$_3$Ir$_4$Sn$_{13}$. 

\begin{figure}[h] 
	\includegraphics[width=\linewidth]{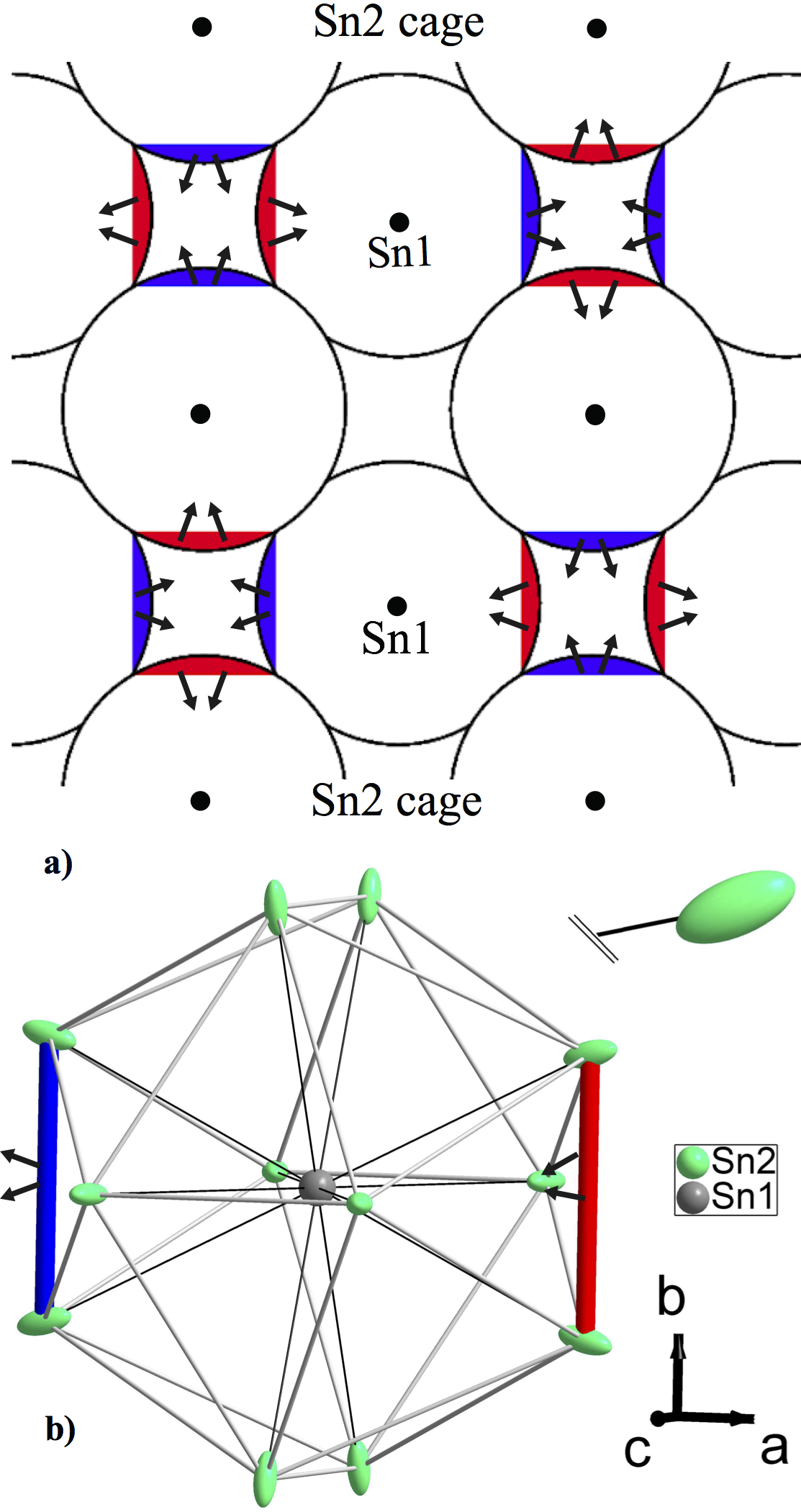}
	\caption{\label{DFT}
		(Color online) \textbf{a)} Illustration of the breathing of Sn$_{12}$ icosahedra in (Ca,~Sr)$_3$Ir$_4$Sn$_{13}$. The arrows represent the atomic replacements of the Sn2 atoms on the surface of the icosahedra. \textbf{b)} Thermal ellipsoids of the Sn1 and Sn2 atoms at $T$~=~50~K with the suggested 'breathing' mode. Inset: Orientation of the thermal ellipsoid of Sn2 and of the Sn1~-~Sn2 bond.}
\end{figure}

The soft phonon mode at (1/2,~1/2,~0) has the largest amplitude on the Sn$_{12}$ (icosahedral) polyhedron formed by the Sn2 atoms. The Sn$_{12}$ icosahedra form a sublattice with a 'BCC' structure (see Fig. \ref{crystalstructure}). Fig. \ref{DFT} a illustrates the Sn2 displacements associated to this mode. We conclude therefore that the $\vec{q}$~=~(1/2,~1/2,~0) mode associated with the Sn2 atoms is related to the phase transitions in Ca$_3$Ir$_4$Sn$_{13}$ and Sr$_3$Ir$_4$Sn$_{13}$. This scenario is supported by the shape and the orientation of the thermal ellipsoids for $T$~$>$~$T^*$ obtained from neutron diffraction (see below).

There are further optical modes involving the Sn atoms, but they do not give rise to lattice instabilities. Naively, one might have expected that the Sn1 atoms  can rattle inside their cages in a cooperative mode. However, the Sn2 atoms of the cage have higher vibrational amplitudes than the central Sn1 atom (see Fig. \ref{DFT} a), and therefore the Sn1 atoms are less likely to be substantially involved in the phase transition at $T^*$. 

\subsubsection{Anisotropic mean-square displacements}
The breathing mode suggested by our DFT calculation is reflected in the thermal displacements of the atoms. The thermal ellipsoids of the atoms, $U_{ij}$ ($i$, $j$~=~1, 2, 3), contribute to the refinement via the Debye-Waller factor\cite{Fullprof,inttablecryst} and is restricted by the symmetry of the space group\cite{inttablecryst}.  

The integrated intensities collected on D9 at $T$~=~50~K were refined using the space group $Pm\bar{3}n$ (conventional agreement factor $R_F$~$\sim$~5~\%). In a first step only isotropic mean-squared displacements factors were used. In the last stage of the fit, anisotropic mean-squared displacements were also refined. The resulting thermal ellipsoids were found to be nearly isotropic for all atoms with the exception of Sn2. They display a pronounced cigar shape with the long axis approximately along the Sn1~-~Sn2 bond. This is in perfect agreement with the conclusion of the DFT calculations concerning the atomic displacements associated to the $\vec{q}$~=~(1/2,~1/2,~0) 'breathing' mode (see Fig. \ref{crystalstructure}, Fig. \ref{DFT} b). Similar results were obtained from the refinement of the single crystal x-ray diffraction data above $T^*$ (40 and 280~K).

\section{Summary} 

We investigated the microscopic structural properties of Ca$_3$Ir$_4$Sn$_{13}$ by means of x-ray and neutron diffraction as well as neutron spectroscopy. For temperatures higher than $T^*$ a cubic crystal structure described by the space group $Pm\bar{3}n$ is found, in agreement with earlier reports\cite{Cooper1980}. Single crystal x-ray diffraction reveal a structural modulation with propagation vector of $\vec{q}$~=~(1/2,~1/2,~0) for $T$~$<$~$T^*$. However, it is not possible to distinguish a description based on a tetragonal unit cell and the $I$-centered double cubic unit cell recently proposed\cite{Klintberg2012} in the low-temperature phase of Sr$_3$Ir$_4$Sn$_{13}$. In the first scenario three equivalent tetragonal domains are formed at $T$~$<$ ~$T^*$.
The temperature dependence of the satellite $\vec{Q}$~=~(7/2,~7/2,~1) in the low-temperature phase reveals a second-order phase transition with a critical temperature $T^*$~=~38.5(2)~K and a critical exponent that is well understood in terms of mean-field theory. 
The inelastic neutron data evidence the softening of a low-energy phonon at $\vec{Q}$~=~(7/2,~7/2,~1) above $T^*$. The temperature dependence of this phonon reveals a displacive, second-order phase transition described by mean-field theory. Single crystal diffraction results and a theoretical search for this soft mode suggest the freezing of a 'breathing' mode of the Sn2 atoms associated with a propagation vector $\vec{q}$~=~(1/2,~1/2,~0). 

\begin{acknowledgments} 
We thank A.~Bosak for the preparation of the x-ray single-crystalline samples and F. M. Grosche for useful inputs. We acknowledge the European Synchrotron Radiation Facility, the Institut Laue-Langevin, the Swiss Light Source and the Neutron Source SINQ at the Paul Scherrer Institut for the allocated beam time. This work was supported by the Swiss NSF under Grant No. 200021\_147071, 200021-122054, 200020-140345, Fellowship No.  P2EZP2\_148737 and the NCCR MaNEP. 
\end{acknowledgments}

\bibliography{Ca3Ir4Sn13_01} 

\end{document}